# Quality factor of plasmonic monopartite and bipartite surface lattice resonances

Joshua T.Y. Tse and H.C. Ong[a]

Department of Physics, The Chinese University of Hong Kong, Shatin, Hong Kong, People's Republic of China

Surface lattice resonance (SLR) is the collective excitation of nanoparticle resonances arising from the hybridization between localized surface plasmons (LSPs) and propagating Rayleigh anomalies (RAs). When comparing with the corresponding LSPs, SLRs exhibit much higher quality factor. In fact, as the quality factor depends on the constituting resonances and their hybridization, how one can parametrize it in an analytic form is an important issue. We have studied the SLRs arising from 2D Au monopartite nanoparticle arrays by angle- and polarization-resolved reflectivity spectroscopy, temporal coupled mode theory (CMT) and finite-difference time-domain (FDTD) simulation. The scattering matrix of the SLRs is formulated, revealing the importance of the spectral detuning and the interaction strengths between the LSP and the RAs in governing the quality factor. We then extend the CMT approach to study bipartite arrays where nanoparticle dimer is employed and find the coupling between two LSPs plays a major role in further boosting the quality factor. Specifically, the coupling takes part in controlling the detuning factor as well as determining whether the coupled bright or dark mode is hybridized with the RAs. The dark mode hybridization can strongly enhance the quality factor which is otherwise not possible in the monopartite counterparts.

[a] Email: hcong@phy.cuhk.edu.hk

## I. INTRODUCTION

Periodic plasmonic systems have gained worldwide attention due to their ability to localize electromagnetic fields beyond diffraction limit, thus providing extraordinary field enhancement [1-5]. In general, two classes of periodic plasmonic systems are available and they are nanohole and nanoparticle arrays [1-5]. While nanohole array is flat metallic surface perforated with lattice of subwavelength holes, nanoparticle array is a lattice of nanoparticles placed on dielectric substrate [1-8]. Other than their complementary system configurations, the plasmonic resonances arising from nanohole and nanoparticle arrays are also different [1-3,9-13]. Nanohole arrays support Bloch-like surface plasmon polaritons (SPPs) which propagate on flat metal surface whereas nanoparticle arrays make use of the coupling between the diffractive Rayleigh anomalies (RAs) and the localized surface plasmons (LSPs) to support the so-called surface lattice resonance (SLR) [1-3,9-13].

Because nanohole and nanoparticle arrays support different resonances, they are expected to exhibit different dispersion relations and interactions with far-fields. As an example, for a two-dimensional (2D) square lattice, the nondegenerate (-1,0) resonance propagates along the Γ-X direction. For Bloch-like SPPs, since both the transverse and longitudinal electric fields of SPPs lies in the plane defined by the propagation direction, only *p*-polarized light can be used for excitation, and the far-field radiation damping remains as *p*-polarized [14]. On the other hand, for SLR, the radiations from the LSPs should align with the propagation direction defined by the (-1,0) RA in order to couple all LSPs into a collective resonance [10]. Therefore, *s*-polarized light, instead of *p*-, is required to excite the (-1,0) SLR [10]. In other words, if one compares the dispersion relations between nanohole and nanoparticle arrays under *p*- and *s*-polarizations, two are completely reversed, indicating the excitation and decay mechanisms for two systems are completely different.

In addition, because of the low dissipative RAs, SLRs enjoy a high quality (Q) factor, which is one to two orders of magnitude higher than LSPs [15-17]. In fact, some studies have reported the Q factor can go beyond 2000 at near infrared regime [18]. These studies are usually conducted on monopartite systems where only one single nanoparticle is present in a unit cell [15-19]. Remarkably, for some of the bipartite systems, in which the unit cell has two nanoparticles, they can exhibit much higher Q factor than the monopartite counterparts [20-23]. Such high Q factors are found to be strongly dependent on the relative position and orientation between two nanoparticles [21]. As the Q factor is determined by how the SLRs decay, the study of the interactions between LSPs and RAs deserves further attention. More importantly, it is desired if the Q factor can be analytically formulated so that one can tailor it at will.

Here, we have studied the Q factor of SLRs from 2D Au nanodisk arrays. We first use angle- and polarization-resolved reflectivity spectroscopy to map the dispersion relations of the monopartite arrays and reveal a variety of resonances including LSP and SLRs. We then formulate the scattering matrix for the SLRs based on temporal coupled mode theory (CMT) to understand the dependence of the decay rates on the properties of the LSP and RAs as well their interactions. From CMT, we see the Q factor of SLRs depends strongly on the spectral detuning and the coupling strength between the LSP and RAs. It is inversely proportional to the coupling strength and at the same time scales quadratically with the detuning. We further extend the CMT to the bipartite systems and find the far-field interaction between the LSPs plays an important role in determining whether a dark or a bright mode is coupled with the RAs. While low Q factor is observed from the bright mode coupling, the dark mode coupling can result in extremely high Q factor that surpasses the monopartite systems. We also demonstrate varying the geometry of the dimer can effectively control the far-field interaction and thus the resulting Q factor.

## II. METHOD

We have prepared 2D square lattice Au cylindrical nanodisk arrays on glass substrate by electron-beam lithography. The scanning electron microscopy plane-view image of one of the arrays is illustrated in the inset of Fig. 1(a), showing the lattice has period $P = 400$ nm, and the basis has height $H$ and radius $R = 50$ and 80 nm. After sample preparation, the sample is then transferred to a homebuilt optical microscope designed for polarization- and angle-resolved reflectivity spectroscopy [24] (see Fig. 1(a)). The schematic of the system is displayed in Fig. 1(b). Briefly, a supercontinuum generation white light laser output from a nonlinear photonic crystal fiber is first coupled to a collimator that expands the beam size to 500 mm$^2$ and then focused onto the back focal plane (BFP) of a 100X oil-immersion objective lens. The light exiting the objective lens will be collimated again at angle $\theta$ defined by $d = f\sin\theta$, where $f$ is the focal length of the objective lens and $d$ is the displacement between the focused beam spot on the BFP and the optical axis of the objective lens [24,25]. By placing the entire illumination optics on a motorized translation stage so that we can translate the beam across the BFP, the incident polar angle $\theta$ can be varied from 0° to 60° with angular resolution as small as 0.1°, provided the numerical aperture (NA) of the objective lens is 1.3. The sample is placed on a motorized rotation stage so that the azimuthal angle $\varphi$, defined with respect to the Γ-X direction, can be varied as well. Because of the refractive index matching oil, the nanodisk array is immersed in a homogeneous environment. The specular reflection from the sample is then collected by the same objective lens and is routed to a spectrometer coupled with a CCD detector for spectroscopy. Finally, the polarization can be controlled by placing a pair of polarizer and analyzer in the incident and detection paths.

## III. EXPERIMENTAL RESULTS

Fig. 2(a) & (b) show the *s*- and *p*-polarized $\theta$-resolved reflectivity mappings of the sample taken along the Γ-X direction, i.e. $\varphi = 0°$, whereas Fig. 2(c) & (d) show the corresponding $\varphi$-resolved reflection mappings with $\theta = 30°$. One can see the dispersive and non-dispersive modes, which are identified as RAs, LSP and SLRs. For the dispersive modes, the RAs, as indicated by the white dash lines in Fig. 2(a) - (d), are calculated by the phase matching equation given as [1-3]:

$$\left(\frac{2\pi}{\lambda}\right)^2 = \left(\frac{2\pi}{\lambda}\sin\theta\cos\varphi + \frac{n2\pi}{P}\right)^2 + \left(\frac{2\pi}{\lambda}\sin\theta\sin\varphi + \frac{m2\pi}{P}\right)^2, \quad (1)$$

where $\lambda$ is the wavelength and ($n,m$) is the mode order of the RA. We see the dispersive modes match very well with the (-1,0) and (0,±1) RA. On the other hand, the nondispersive mode at $\lambda = $ 790 nm (see Fig. 2(b)) is attributed to the LSP arising from the nanodisks. In fact, by performing FDTD on a single nanodisk, the resonance is found to be 779 nm, which agrees with the experimental result [26]. The high reflection peaks adjacent to the RAs are assigned as SLRs. For example, in the *s*-polarized $\theta$-resolved mapping in Fig. 2(a), the LSP interacts with (-1,0) RA to form (-1,0) SLR along the Γ-X direction. Likewise, in the $\varphi$-resolved mapping in Fig. 2(c), we see the LSP mode couples with the (-1,0) and (0,-1) RAs, lifting the LSP to ~ 700 nm at $\varphi = 0°$ – 20° and 70° – 90° and at the same time forming (-1,0) and (0,-1) SLRs. However, for the *p*-polarized counterparts in Fig. 2(b) taken along the Γ-X direction, as the radiation fields from the LSP are no longer aligned with the RAs, no SLRs are observed [10]. Other than the nondegenerate (-1,0) and (0,-1) SLRs, we also see the degenerate (0,±1) SLRs can be excited by both *s*- and *p*-polarized lights in Fig. 2(a) & (b). The (0,±1) RAs both can collectively couple with the fields radiated from *s*- and *p*-excited LSP. In addition, in analogy to the Bloch-like SPPs, the (0,±1) SLRs excited by different polarizations possess different field symmetries with respect to the incident plane [27]. While *p*-excited SPPs are symmetric in accordance with the incident wave, *s*-excited SLRs are anti-symmetric [10]. A few SLR spectra are illustrated in Fig. 3(a) - (d), showing the SLRs are Fano-like [8]. For now, we will focus on formulating the scattering matrix for the nondegenerate (-1,0) SLR where its resonant wavelength is longer than that of the LSP.

## IV. TEMPORAL COUPLED MODE THEORY

We study the properties of SLR within the framework of temporal CMT [28,29]. Since SLR arises from the interactions between a LSP and two *s*- and *p*-RAs, their dynamics can be written as:

$$\frac{d}{dt}\begin{bmatrix} a_1 \\ a_2 \\ a_3 \end{bmatrix} = i\begin{bmatrix} \tilde{\omega}_1 & \Omega_{12} & \Omega_{13} \\ \Omega_{12} & \tilde{\omega}_2 & 0 \\ \Omega_{13} & 0 & \tilde{\omega}_3 \end{bmatrix}\begin{bmatrix} a_1 \\ a_2 \\ a_3 \end{bmatrix} + K|S_+\rangle, \tag{2}$$

where $\tilde{\omega}_{1-3} = \omega_{1-3} + i\Gamma_{1-3}/2$, and $a_{1-3}$, $\omega_{1-3}$, and $\Gamma_{1-3}$ are the mode amplitudes, resonant frequencies, and total decay rates of the LSP, *s*- and *p*-RAs. $\Omega_{12}$ and $\Omega_{13}$ are the coupling constants between the LSP and *s*-RA and the LSP and *p*-RA, respectively, and $K$ and $|S_+\rangle$ are the complex in-coupling matrix and the incident power amplitude vector [30]. We expect the *s*- and *p*-RAs are degenerate but do not interact due to their orthogonality. Both the *s*- and *p*-RAs are dark modes in which $\Gamma_2$ and $\Gamma_3$ are identical and small compared to $\omega_{1-3}$ so that they are not driven by the incident light [8]. In addition, the LSP radiates over the entire space in a homogeneous environment with the total decay rate defined as $\Gamma_1 = \Gamma_{1,abs} + \int \Gamma_{1,rad}^{\theta,\varphi} d\Omega$, where $\Gamma_{1,abs}$ is the absorption decay rate and $\Gamma_{1,rad}^{\theta,\varphi}$ is the radiative decay rate at $\theta$ and $\varphi$, and $d\Omega = \sin\theta d\theta d\varphi$ is the differential solid angle [31]. By solving the determinant of the homogeneous part of Eq. (2), we have $\tilde{\omega}'_{1,2} = \frac{\tilde{\omega}_1 + \tilde{\omega}_2}{2} \pm \sqrt{\left(\frac{\tilde{\omega}_1 - \tilde{\omega}_2}{2}\right)^2 + \Omega_{12}^2 + \Omega_{13}^2}$ and $\tilde{\omega}'_3 = \tilde{\omega}_2$ for the eigenfrequencies and $a'_{1,2} = \begin{bmatrix} \tilde{\omega}_2 - \tilde{\omega}'_{1,2} & -\Omega_{12} & -\Omega_{13} \end{bmatrix}^T$ and $a'_3 = \begin{bmatrix} 0 & \Omega_{13} & -\Omega_{12} \end{bmatrix}^T$ for the eigenvectors. In fact, as illustrated in the Supplementary Information [26], Eq. (2) is diagonalizable so that after transformation we can treat the nondegenerate SLR as a single mode [32]. If $\omega_1 > \omega_{2,3}$ for our case, the (-1,0) SLR has mode amplitude *a*, following:

$$\frac{da}{dt} = i\omega_o a - \frac{\Gamma_{tot}}{2}a + \sqrt{\frac{\Gamma_{rad}}{2}}\langle\kappa^*|s_+\rangle, \tag{3}$$

where $\omega_0$ is the resonant frequency and $\Gamma_{tot}$ is the total decay rate of SLR, which is the summation of the absorption and radiative decay rates, $\Gamma_{abs}$ and $\Gamma_{rad}$ [32]. Since the SLR supports only the specular transmission and reflection, the incident wave vector $|s_+\rangle = \begin{bmatrix} s^R_{+,s} & s^R_{+,p} & s^T_{+,s} & s^T_{+,p} \end{bmatrix}^T$ and the in-coupling constant vector $|\kappa\rangle = \begin{bmatrix} e^{i\delta_s}\cos\alpha & e^{i\delta_p}\sin\alpha & e^{i\delta_s}\cos\alpha & e^{i\delta_p}\sin\alpha \end{bmatrix}^T$, where the subscripts $s/p$ define the polarizations, the superscripts $R/T$ are the reflection and transmission sides, $\alpha$ is the best incident polarization angle, and $\delta_{s/p}$ are the in-coupling phase shifts [30]. The incident wave vector is normalized as $\langle s_+|s_+\rangle = 1$. Under the conservation of energy and time reversal symmetry, the outgoing fields can be expressed as [28-30]:

$$|s_-\rangle = C|s_+\rangle + \sqrt{\frac{\Gamma_{rad}}{2}}a|\kappa\rangle, \tag{4}$$

where $C$ is the direct scattering matrix given as $\begin{bmatrix} t_{o,s} & 0 & r_{o,s} & 0 \\ 0 & t_{o,p} & 0 & r_{o,p} \\ r_{o,s} & 0 & t_{o,s} & 0 \\ 0 & r_{o,p} & 0 & t_{o,p} \end{bmatrix}$ in which $t_{o,s/p}$ and $r_{o,s/p}$ are the direct transmission and reflection coefficients and there is no polarization conversion arising from direct scattering. As shown by Fan et al. [29,33], the direct scattering matrix should be unitary, i.e. $C^\dagger C = I$, the in- and out-coupling constants are normalized as $\langle\kappa|\kappa\rangle = 2$, and they together should satisfy $C|\kappa^*\rangle = -|\kappa\rangle$.

To obtain the far-fields of the SLR under the excitation from the reflection side, we solve Eq. (3) to have $a = \sqrt{\frac{\Gamma_{rad}}{2}}\frac{\langle\kappa^*|s_+\rangle}{i(\omega-\omega_o)+\Gamma_{tot}/2}$. With the aid of Eq. (4), we can now write the scattering matrix $S$ defined as $|s_-\rangle = S|s_+\rangle$ as:

$$S = C + \frac{\Gamma_{rad}}{2}\frac{|\kappa\rangle\langle\kappa^*|}{i(\omega-\omega_o)+\Gamma_{tot}/2}. \tag{5}$$

Once the scattering matrix is available, we formulate the $s$- and $p$-transmission and reflection spectral profiles. The $s$- and $p$-polarized transmissions can be written as:

$$T_s = \left| t_{o,s} + \frac{\Gamma_{rad} e^{2i\delta_s} \cos^2 \alpha/2}{i(\omega - \omega_o) + \Gamma_{tot}/2} \right|^2 \text{ and } T_p = \left| t_{o,p} + \frac{\Gamma_{rad} e^{2i\delta_p} \sin^2 \alpha/2}{i(\omega - \omega_o) + \Gamma_{tot}/2} \right|^2 \text{ under } |s_+\rangle = \begin{bmatrix} 1 & 0 & 0 & 0 \end{bmatrix}^T \text{ and}$$

$|s_+\rangle = \begin{bmatrix} 0 & 1 & 0 & 0 \end{bmatrix}^T$. On the other hand, the $s$- and $p$-polarized reflections are

$$R_s = \left| r_{o,s} + \frac{\Gamma_{rad} e^{2i\delta_s} \cos^2 \alpha/2}{i(\omega - \omega_o) + \Gamma_{tot}/2} \right|^2 \text{ and } R_p = \left| r_{o,p} + \frac{\Gamma_{rad} e^{2i\delta_p} \sin^2 \alpha/2}{i(\omega - \omega_o) + \Gamma_{tot}/2} \right|^2,$$ respectively. They show the

far-field spectra consists of two parts and they are direct scattering and radiation damping from the SLR mode, which follows a Lorentzian line shape with the linewidth defined by $\Gamma_{tot}$. In the homogeneous environment, although the direct reflection is weak, it interferes with the radiation damping from SLR, yielding a Fano peak-like spectrum [8,34], which is consistent with our experimental results.

## V. STUDIES OF DECAY MECHANISM AND Q FACTOR

We then study the decay mechanism of the SLR by FDTD and experiment. We first verify the CMT by FDTD for determining $\Gamma_{rad}$ and $\Gamma_{abs}$. The simulations are performed on a unit cell with $P = 400$ nm, $H = 50$ nm and $R = 70$ nm. The nanodisks are made of Au with the complex dielectric constants obtained from Ref. [35] and they are immersed in a homogeneous medium with refractive index = 1.5. Bloch boundary condition is applied to four sides of the unit cell while perfectly matched layers are set at the top and bottom. Several simulated transmission and reflection spectra taken along the Γ-X direction and at incident polar angle $\theta = 30°$ and azimuthal angle $\varphi = 6°$-$30°$ under $s$- and $p$-polarizations are shown in Fig. 4(a) - (h). The spectra are then fitted with $T_{s/p}$ and $R_{s/p}$ to determine $\Gamma_{rad}$, $\Gamma_{abs}$, and $\alpha$, and the results are shown in Fig. 5(a) & (b) as a function of resonant wavelength and $\varphi$. The best fits are also shown as the dash lines in the figures, demonstrating good fits.

To verify the fitted values of $\Gamma_{rad}$ and $\Gamma_{abs}$, we have conducted independent time-domain simulations on the same system [26]. The time-domain results are overlaid in Fig. 5(a) and they agree very well with the CMT results. On the other hand, for the confirmation of $\alpha$, for each excitation angle, we calculate the absorption of the SLR mode as a function of the incident polarization angle $\gamma$ defined with respect to the $s$-polarization and identify the best $\gamma$ that yields the strongest absorption [30]. The results are provided in Fig. 5(b), showing the error is within 5%.

Therefore, we conclude the scattering matrix is adequate to describe the angle- and polarization-dependent transmission and reflection spectra of (-1,0) SLR.

Our formalism also reveals the decay mechanism and the Q factor of the SLR. First, if both *s*- and *p*-RAs are considered as almost completely dark, far-field interactions between them and the LSPs are forbidden. As only near-field interactions prevail, both $\Omega_{12}$ and $\Omega_{13}$ are real values [36]. The total decay rate $\Gamma_{tot}$ is $\frac{\Gamma_1+\Gamma_2}{2} - \text{Im}\left[\sqrt{\left(\omega_1-\omega_2+\frac{i}{2}(\Gamma_1-\Gamma_2)\right)^2 + 4(\Omega_{12}^2+\Omega_{13}^2)}\right]$, which indicates its dependence on the properties of LSP and RAs. By expanding the imaginary part of the square root, we find $\Gamma_{tot}$ can be approximated as $\Gamma_2 + \frac{\Gamma_1}{4}\left(\frac{(\omega_1-\omega_2)^2+(\Gamma_1/2)^2}{4(\Omega_{12}^2+\Omega_{13}^2)}+1\right)^{-1}$, showing it scales with the decay rates of *s*-RA and LSP, in which the rate of LSP is further modified by a factor that depends on the interplay between the LSP and RA spectral detuning, i.e. $\omega_1-\omega_2$, and the coupling constants $\Omega_{12}$ and $\Omega_{13}$ [26]. In particular, for a system with fixed $\Omega_{12}$ and $\Omega_{13}$, $\Gamma_{tot}$ decreases quadratically with the spectral detuning. We plot $1/\Gamma_{tot}$ against $(\omega_1-\omega_2)^2$ in Fig. 5(c) and it clearly shows a linear dependence. Likewise, we find the Q factor, which is defined as $\omega_o/\Gamma_{tot}$, can be approximated as:

$$\frac{4}{\Gamma_1}\left(\omega_2\left(\frac{(\omega_1-\omega_2)^2+(\Gamma_1/2)^2}{4(\Omega_{12}^2+\Omega_{13}^2)}+2\right)-\omega_1\right). \tag{6}$$

A linear dependence is shown in Fig. 5(d) where the Q factor is plotted as a function of $\omega_2(\omega_1-\omega_2)^2$. On the other hand, $\Gamma_{rad}$ in Fig. 5(a) shows a $\lambda^{-n}$ dependence, where *n* is fitted to be 8.7, showing the radiation damping of the SLR follows quadrupole like Mie scattering [37,38]. Considering the size of the nanodisk is ~1/5 of the resonant wavelength, higher order, instead of dipolar, Mie scattering thus is expected.

We are now in the position of applying the CMT to fit our experimental spectra to determine $\Gamma_{rad}$, $\Gamma_{abs}$ and $\alpha$. We fit the reflectivity spectra in Fig. 3(a) - (d) by using $R_{s/p}$ and the best fits are displayed as the dash lines. The results together with $\Gamma_{tot}$ are plotted in Fig. 6(a) - (c). From the figures, they show the trends of $\Gamma_{rad}$, $\Gamma_{abs}$, $\alpha$, and $\Gamma_{tot}$ follow well as the simulations in Fig.

5(a) - (c). In particular, $\Gamma_{rad}$ in Fig. 6(a) shows a similar $\lambda$ dependence in which $n$ is determined to be 13.9 whereas $1/\Gamma_{tot}$ in Fig. 6(c) displays a linear dependence on $(\omega_1 - \omega_2)^2$. Fig. 6(d) verifies the linear dependence of the Q factor on $\omega_2(\omega_1 - \omega_2)^2$. On the other hand, as $\alpha$ is interpreted as the angle where the overlap between the projection of the incident polarization and the SLR transverse near-field is maximal so that the energy transfer between the far- and near-fields is optimal, we can expect $\tan\alpha = \cos\theta \tan\rho$, where $\rho$ is the propagation direction of SLR defined with respect to the incident plane [24,30]. The results are then overlaid in Fig. 6(c), in consistent with the experiment.

## VI. BIPARTITE SYSTEMS

After studying the monopartite systems, we then extend the CMT approach to bipartite nanoparticle arrays. For dimer arrays shown in the inset of Fig. 7(a), they support two identical LSPs that couple differently in the $x$- and $y$-directions. However, only when the radiation fields align with the RA yields SLRs. If the RA propagating in the $x$-direction, we simply neglect the coupling of the LSPs in the $x$-direction. We therefore extend Eq. (2) to bipartite system as:

$$\frac{d}{dt}\begin{bmatrix} a_1 \\ a_2 \\ a_3 \\ a_4 \end{bmatrix} = i \begin{bmatrix} \tilde{\omega}_1 & \tilde{\Omega}_{12} & \Omega_{13} & \Omega_{14} \\ \tilde{\Omega}_{12} & \tilde{\omega}_1 & \Omega_{13} & \Omega_{14} \\ \Omega_{13} & \Omega_{13} & \tilde{\omega}_3 & 0 \\ \Omega_{14} & \Omega_{14} & 0 & \tilde{\omega}_3 \end{bmatrix} \begin{bmatrix} a_1 \\ a_2 \\ a_3 \\ a_4 \end{bmatrix} + K|S_+\rangle \quad (7)$$

where $a_{1,2}$ are the mode amplitudes of the LSPs with resonant frequency $\tilde{\omega}_1$, $a_{3,4}$ are the mode amplitudes of the $s$- and $p$-RAs that have $\tilde{\omega}_3$, $\tilde{\Omega}_{12}$ is the coupling constant between two LSPs, and $\Omega_{13}$ and $\Omega_{14}$ are the coupling constants between the LSPs and the RAs. In addition, unlike $\Omega_{13}$ and $\Omega_{14}$ which are real constants, $\tilde{\Omega}_{12} = \Omega'_{12} + i\Omega''_{12}$ is complex so that both near- and far-field interactions take place considering the nanoparticles are highly radiative [31,36,39-40]. In particular, $\Omega''_{12}$ is an anti-Hermitian coupling constant that associates with primarily far-field interaction [41]. We diagonalize the homogeneous part of Eq. (7) and find the eigenfrequencies to be $\tilde{\omega}'_1 = \tilde{\omega}_1 - \tilde{\Omega}_{12}$, $\tilde{\omega}'_2 = \tilde{\omega}_3$ and $\tilde{\omega}'_{3,4} = \frac{\tilde{\omega}_1 + \tilde{\Omega}_{12} + \tilde{\omega}_3}{2} \pm \sqrt{\left(\frac{\tilde{\omega}_1 + \tilde{\Omega}_{12} - \tilde{\omega}_3}{2}\right)^2 + 2(\Omega_{13}^2 + \Omega_{14}^2)}$ [26]. At the same time, the corresponding eigenvectors are $a'_1 = \begin{bmatrix} -1 & 1 & 0 & 0 \end{bmatrix}^T$, $a_2' = \begin{bmatrix} 0 & 0 & \Omega_{14} & -\Omega_{13} \end{bmatrix}^T$ and $a_{3,4}' = \begin{bmatrix} \tilde{\omega}_3 - \tilde{\omega}_{3,4}' & \tilde{\omega}_3 - \tilde{\omega}_{3,4}' & -2\Omega_{13} & -2\Omega_{14} \end{bmatrix}^T$. We see, after the transformation, $\tilde{\omega}_{1,2}'$ remain as nondispersive LSP and dispersive RA like. On the other hand, for $\omega_1 > \omega_3$, we can identify $a_4'$ as the (-1,0) SLR in which its resonant frequency is

$$\omega_4' = \frac{\omega_1 + \Omega_{12}' + \omega_3}{2} - \text{Re}\left[\sqrt{\left(\frac{\tilde{\omega}_1 + \tilde{\Omega}_{12} - \tilde{\omega}_3}{2}\right)^2 + 2\left(\Omega_{13}^2 + \Omega_{14}^2\right)}\right]$$

and total decay rate is

$$\Gamma_4' = \frac{\Gamma_1 + 2\Omega_{12}'' + \Gamma_3}{2} - \text{Im}\left[\sqrt{\left(\tilde{\omega}_1 + \tilde{\Omega}_{12} - \tilde{\omega}_3\right)^2 + 8\left(\Omega_{13}^2 + \Omega_{14}^2\right)}\right].$$

Following the Supplementary Information [26], we approximate the total decay rate of the SLR as

$$\Gamma_4' \approx \Gamma_3 + \frac{\Gamma_1 + 2\Omega_{12}''}{4}\left(\frac{\left(\omega_1 + \Omega_{12}' - \omega_3\right)^2 + \left(\Gamma_1/2 + \Omega_{12}''\right)^2}{8\left(\Omega_{13}^2 + \Omega_{14}^2\right)} + 1\right)^{-1},$$

so that the Q factor, i.e. $\omega_4'/\Gamma_4'$, can be approximated as:

$$\frac{4}{\Gamma_1 + 2\Omega_{12}''}\left(\omega_3\left(\frac{\left(\omega_1 + \Omega_{12}' - \omega_3\right)^2 + \left(\Gamma_1/2 + \Omega_{12}''\right)^2}{8\left(\Omega_{13}^2 + \Omega_{14}^2\right)} + 2\right) - \left(\omega_1 + \Omega_{12}'\right)\right). \tag{8}$$

Compared with the monopartite Q factor in Eq. (6), we see Eq. (8) contains additional terms given by $\tilde{\Omega}_{12}$ and the bipartite Q factor can be enhanced or reduced depending on the signs and magnitudes of both $\Omega_{12}'$ and $\Omega_{12}''$. Interestingly, when $\Omega_{12}''$ is tuned to be close to $-\Gamma_1/2$, the Q factor tends toward infinity. On the other hand, for a fixed $\tilde{\Omega}_{12}$, the spectral detuning factor now becomes $\left(\omega_1 + \Omega_{12}' - \omega_3\right)^2$. As $\tilde{\Omega}_{12}$ is geometry dependent, one can control it simply by changing the nanoparticle shape and the relative position and orientation between two nanoparticles [42].

We demonstrate the manipulation of $\tilde{\Omega}_{12}$ by FDTD simulations on Au nanodisk dimer arrays following the work of Ref [21]. Two identical nanodisks in the inset of Fig. 7(a) have $H = 50$ nm and $R = 50$ nm and their relative displacement is denoted as $\mathbf{d} = d_x\hat{\mathbf{x}} + d_y\hat{\mathbf{y}}$. The dimer is then placed in a square lattice unit cell with $P = 400$ nm and the bottom nanoparticle is always fixed at $-100\hat{\mathbf{x}} - 100\hat{\mathbf{y}}$ nm from center of unit cell. The arrays again are enclosed in a homogeneous medium with refractive index = 1.5. At $\theta = 24°$ along the Γ-X direction, we calculate the s-polarized transmission and reflection spectra of the arrays with $d_x = 25 - 175$ nm and $d_y = 100$ nm in Fig. 7(a) & (b). The (-1,0) SLRs are identified at ~ 927 – 977 nm. Apparently, consistent with Ref [21], we note the linewidth of the SLRs decreases progressively with increasing $d_x$. To elucidate the dependence of the Q factor on $\tilde{\Omega}_{12}$, we determine $\Omega'_{12}$ and $\Omega''_{12}$ by simulating the $s$-polarized transmission and reflection spectra in Fig. 8(a) & (b) at $\theta = 30°$ where $\left|\tilde{\omega}_1 + \tilde{\Omega}_{12} - \tilde{\omega}_3\right|^2 \gg \left(\Omega^2_{13} + \Omega^2_{14}\right)$ so that $\tilde{\omega}'_{1,3}$ become $\tilde{\omega}_1 \pm \tilde{\Omega}_{12}$, which are also known as the bright and dark modes [42-45], and are spectrally differentiable from the SLRs and RAs. The real and imaginary parts of $\tilde{\Omega}_{12}$ can then be straightforwardly deduced from the differences in the peak positions and the linewidths, which are extracted by fitting the profiles using two Lorentzian peaks. We plot $\Omega'_{12}$ and $\Omega''_{12}$ as a function of $d_x$ in Fig. 8(c), showing while $\Omega'_{12}$ increases gradually with $d_x$ and eventually saturates at $d_x = 100$ nm, $\Omega''_{12}$ decreases monotonically and flips sign at $d_x = 125$ nm. As a result, the bright and dark modes are located at long and short wavelengths when $\Omega''_{12}$ is positive but reversed when $\Omega''_{12}$ flips to negative. More importantly, for negative $\Omega''_{12}$, the dark mode couples with the RAs to form the SLRs, yielding high Q factor. We follow the earlier approach to formulate the scattering matrix as well as $T_s$ and $R_s$ to fit the spectra for determining the Q factors and the results are plotted with $\Omega''_{12}$ in Fig. 7(c). Given the single nanodisk determined by FDTD has $\tilde{\omega}_1 = (2765 + 255i)\,\text{ps}^{-1}$ [26], we calculate $\sqrt{\Omega^2_{13} + \Omega^2_{14}}$ to be $91.8\,\text{ps}^{-1}$ for $d_x = 100$ nm. We then proceed to estimate the Q factor by Eq. (8) and the results are plotted in Fig. 7(c), consistent with the FDTD results. We see in Fig. 7(c) when $\Omega''_{12}$ is approaching $-\Gamma_1/2$, $1/(\Gamma_1 + 2\Omega''_{12})$ becomes more and more dominant over $(\Gamma_1 + 2\Omega''_{12})^2$, strongly enhancing the Q factor to almost 600.

## VII. CONCLUSION

In summary, we have studied the optical properties of SLRs arising from 2D monopartite and bipartite nanodisk arrays by using polarization- and angle-resolved reflectivity spectroscopy, temporal CMT and FDTD. The scattering matrices for the (-1,0) SLRs have been formulated analytically by using CMT and verified by FDTD. The matrices are found to describe the reflection and transmission spectra of the SLRs well. In addition, we show the interplay between the spectral detuning and the coupling constants between the LSPs and RAs plays an important role in governing the decay rate and the Q factor of SLRs. The Q factor increases quadratically with the detuning but depends on the coupling constants in a nontrivial manner. In particular, it is shown in the bipartite systems that controlling the far-field coupling constant between the LSPs can have a dramatic effect on the Q factor.

**VIII. ACKOWLEDGEMENTS**

This research was supported by the Chinese University of Hong Kong through Area of Excellence (AoE/P-02/12) and Innovative Technology Funds (ITS/133/19 and UIM/397).

**Figures**

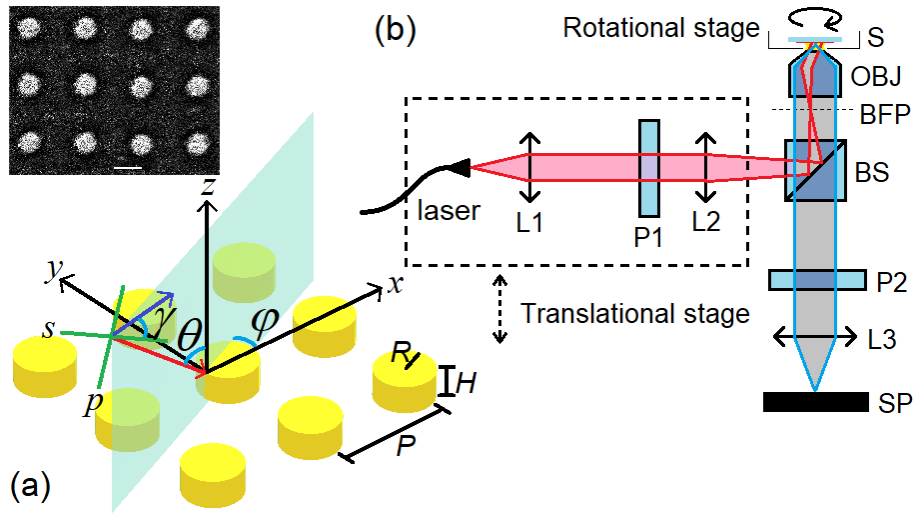

Figure 1. (a) The illustration of the 2D nanodisk array used in experiment and FDTD simulation. The nanodisks with height $H$ and radius $R$ are arranged in a square lattice with period $P$. The Γ-X direction is defined with respect to the $x$-axis. The incident direction, as indicated by the red arrow, is defined by $\theta$ and $\varphi$ with respect to the $z$- and $x$-axes. The incident polarization $\gamma$, as indicated by the blue arrow, is defined with respect to the $s$-polarization. The inset shows the SEM image of the nanodisk array and the scale bar is 200 nm. (b) The schematic of the optical microscope system for angle- and polarization-resolved reflectivity spectroscopy. The supercontinuum generation fiber laser is collimated by a convex lens (L1) and polarized by a polarizer (P1), then is focused by a tube lens (L2) onto the BFP of the oil-immersed objective lens (OBJ). The laser illuminates the sample (S) as a collimated light and the reflected light is captured again by the OBJ. A beamsplitter (BS) routes the reflected light through an analyzer (P2) and is focused by lens (L3) to a CCD-based spectrometer (SP).

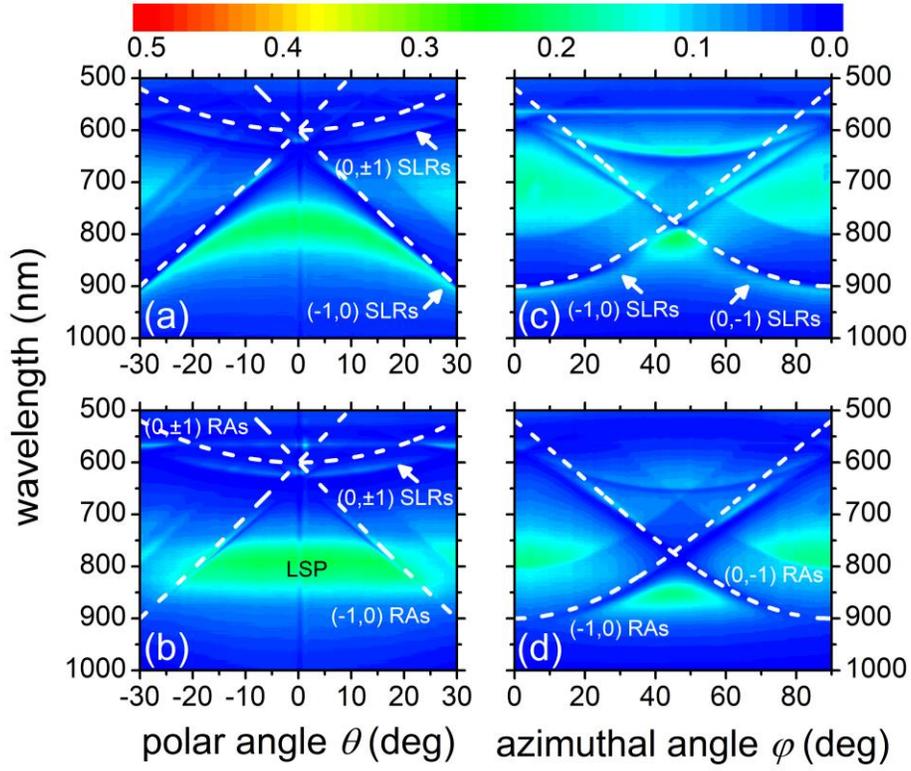

Figure 2. The (a) *s*- and (b) *p*-polarized $\theta$-resolved reflectivity mappings of the array taken along the Γ-X direction, and the (c) *s*- and (d) *p*-polarized $\varphi$-resolved reflectivity mappings at $\theta = 30°$. The white dash lines are the RAs calculated by the phase matching equation, showing (-1,0) and (0,±1) RAs. LSP and different (-1,0) and (0,±1) SLRs are also observed and labelled.

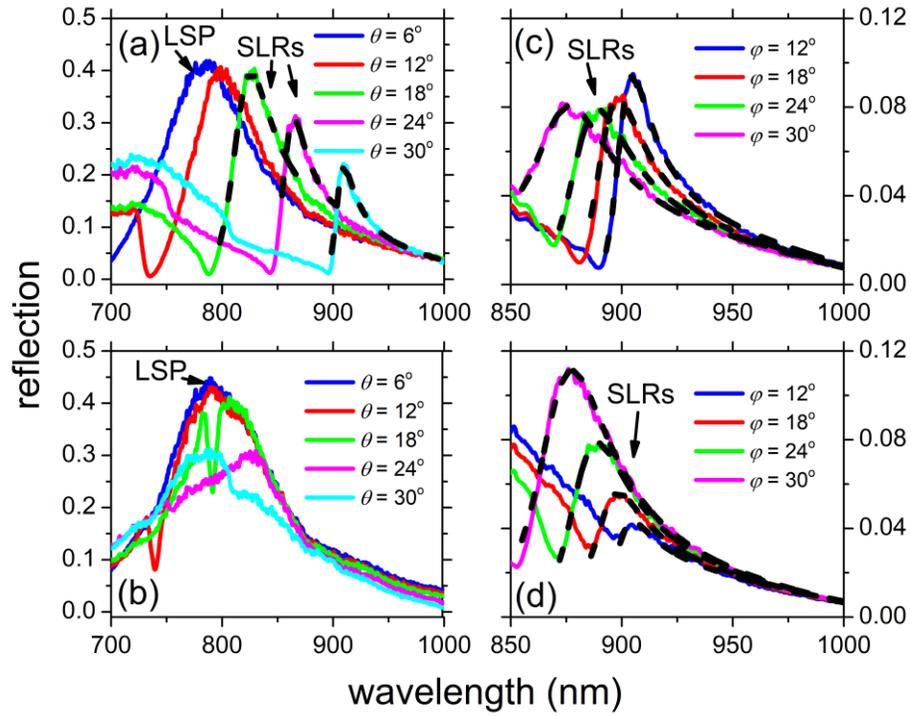

Figure 3. The (a) *s*- and (b) *p*-polarized (-1,0) SLR reflection spectra taken along the Γ-X directions for different $\theta$. The (c) *s*- and (d) *p*-polarized (-1,0) SLR reflection spectra taken at $\theta = 30^\circ$ for different $\varphi$. The black dash lines are the best fits by CMT for determining $\Gamma_{rad}$, $\Gamma_{abs}$, and $\alpha$.

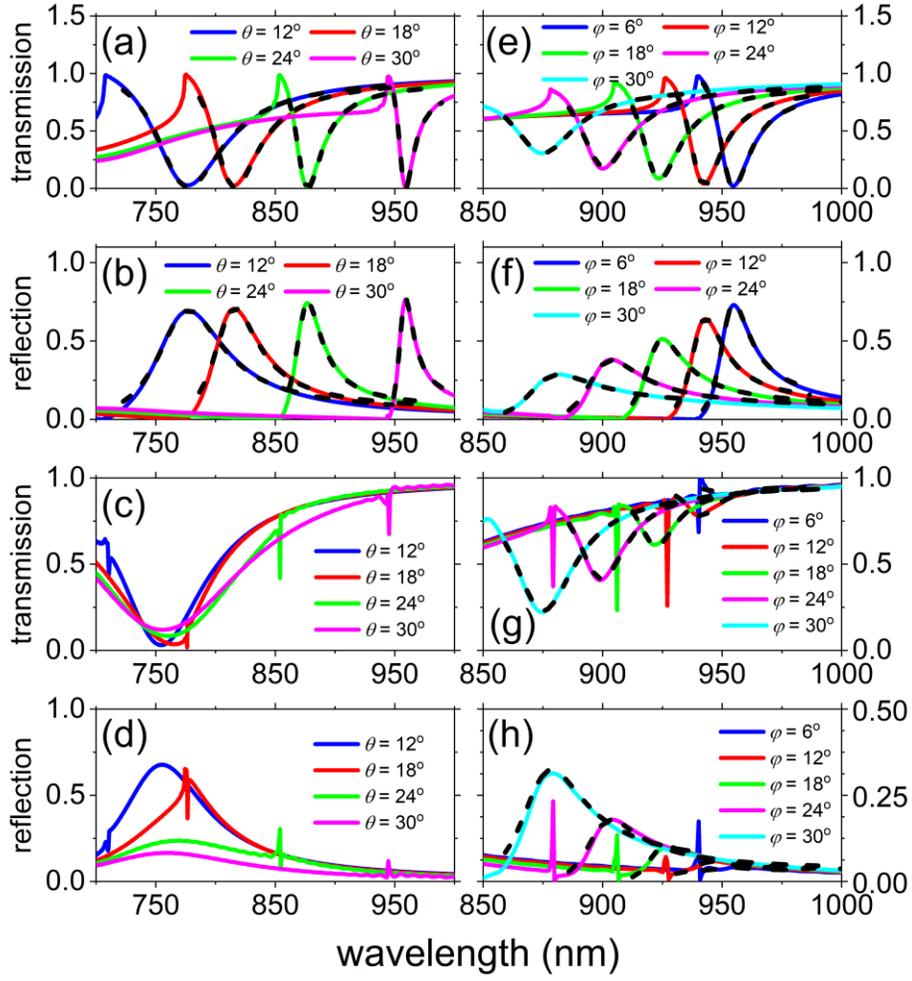

Figure 4. The FDTD simulated (a,b) *s*- and (c,d) *p*-polarized transmission and reflection spectra of (-1,0) SLRs taken along the Γ-X direction for different $\theta$. The (e,f) *s*- and (g,h) *p*-polarized transmission and reflection spectra of (-1,0) SLRs taken at $\theta = 30°$ for different $\varphi$. The black dash lines are the best fits by CMT for determining $\Gamma_{rad}$, $\Gamma_{abs}$, and $\alpha$.

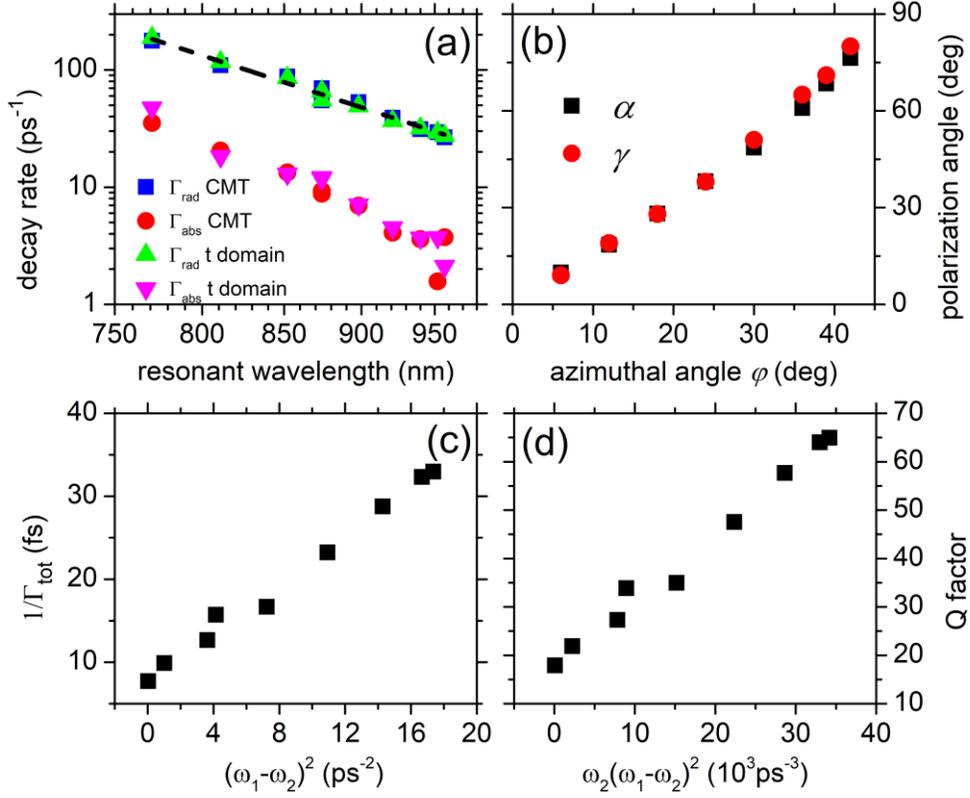

Figure 5. (a) The plots of the FDTD $\Gamma_{rad}$ (blue square) and $\Gamma_{abs}$ (red circle) against the (-1,0) SLR resonant wavelength in log-log scale. The black dash line is the linear best fit, indicating the slope is 8.7 or $\Gamma_{rad} \propto \lambda^{-8.7}$. The time domain simulated $\Gamma_{rad}$ (green up triangle) and $\Gamma_{abs}$ (magenta down triangle) are also plotted for comparison. (b) The plot of the FDTD $\alpha$ (black square) against $\varphi$. The data (red circle) obtained from the $\gamma$ that yields the strongest absorption is also overlaid for comparison. (c) The plot of the FDTD $1/\Gamma_{tot}$ against $(\omega_1 - \omega_2)^2$, exhibiting a linear dependence. (d) The plot of the FDTD Q factor, defined as $\omega_o/\Gamma_{tot}$, against $\omega_2(\omega_1 - \omega_2)^2$, showing a linear dependence.

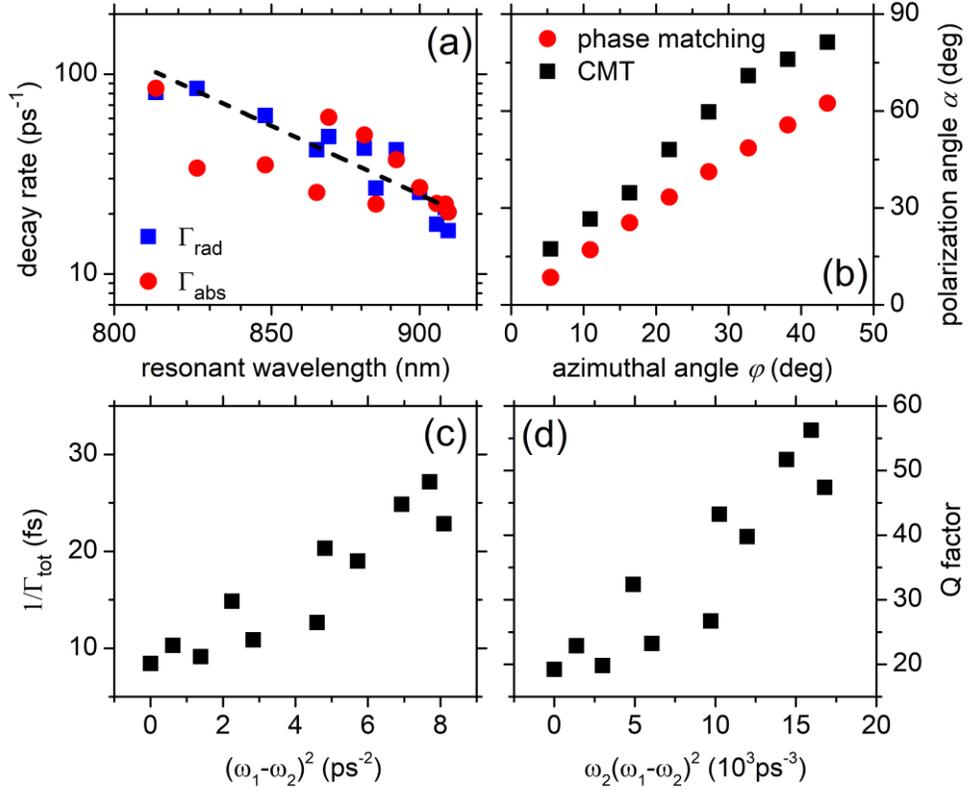

Figure 6. (a) The plots of the experimental $\Gamma_{rad}$ (blue square) and $\Gamma_{abs}$ (red circle) against the (-1,0) SLR resonant wavelength in log-log scale. The dash line is the linear best fit, indicating the slope is 13.9 or $\Gamma_{rad} \propto \lambda^{-13.9}$. (b) The plot of the experimental $\alpha$ (black square) against $\varphi$. The data (red circle) obtained from $\tan\alpha = \cos\theta \tan\rho$ is also overlaid for comparison. (c) The plot of the experimental $1/\Gamma_{tot}$ against $(\omega_1 - \omega_2)^2$, exhibiting a linear dependence. (d) The plot of the experimental Q factor, defined as $\omega_o/\Gamma_{tot}$, against $\omega_2(\omega_1 - \omega_2)^2$, showing a linear dependence. The behaviors of the experimental $\Gamma_{tot}$ and Q factor agree with the FDTD simulated results.

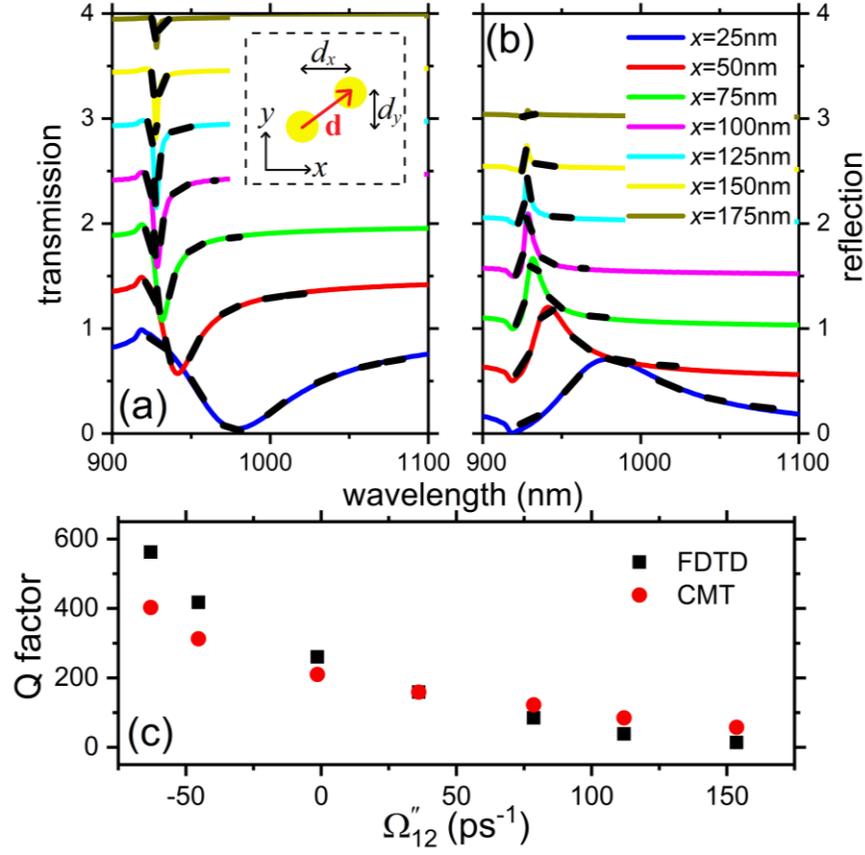

Figure 7. The (a) transmission and (b) reflection spectra of the (-1,0) SLRs taken under *s*-incidence at $\theta = 24°$ along the $\Gamma$-X direction for $d_x = 25 – 175$ nm. The black dash lines are the best fits by CMT for determining the Q factor. The inset shows the unit cell of the bipartite nanoparticle array, characterized by the relative displacement vector $\mathbf{d} = d_x\hat{\mathbf{x}} + d_y\hat{\mathbf{y}}$, as indicated by the red arrow. (c) The plot of the Q factor against $\Omega_{12}''$, showing the analytical CMT and FDTD simulations are consistent.

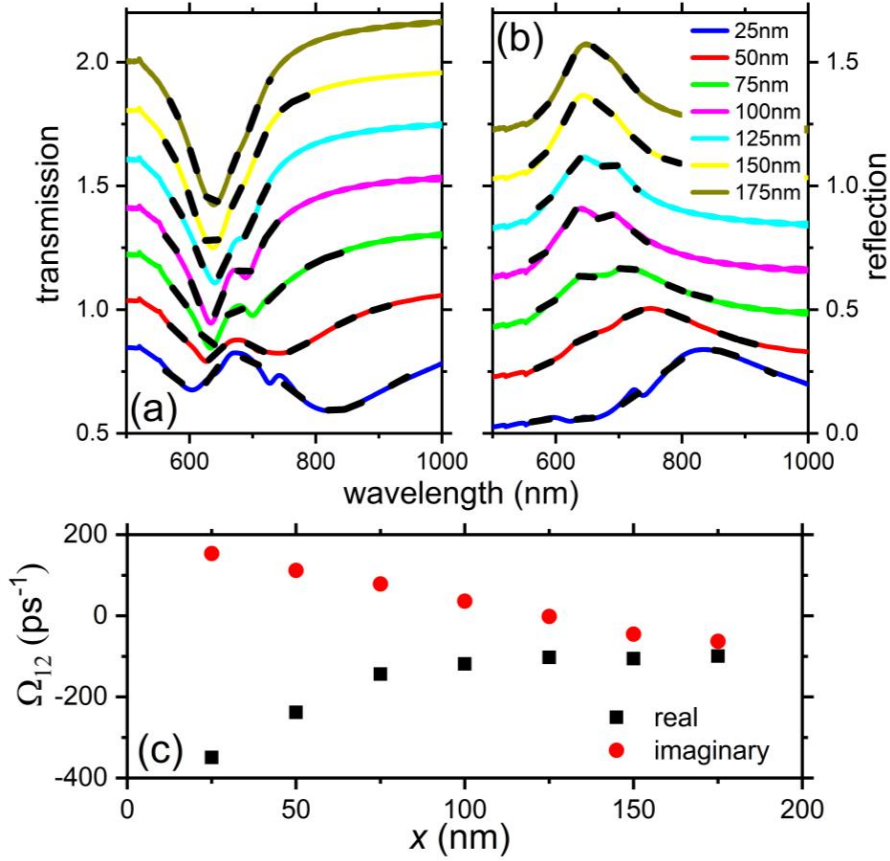

Figure 8. The (a) transmission and (b) reflection spectra of the bright and dark modes taken under $s$-incidence at $\theta = 30°$ along the $\Gamma$-X direction for $d_x = 25 - 175$ nm. The black dash lines are the best fits for determining $\tilde{\Omega}_{12}$. (c) The plot of the real and imaginary component of $\tilde{\Omega}_{12}$ against $d_x$. The imaginary component decreases monotonically from positive to negative, while the real part increases progressively and saturates at $d_x = 100$ nm.

# Supplementary Information

## Quality factor of plasmonic monopartite and bipartite surface lattice resonances

Joshua T.Y. Tse and H.C. Ong

Department of Physics, The Chinese University of Hong Kong, Shatin, Hong Kong, People's Republic of China

### A. FDTD SIMULATION OF SINGLE NANODISK

FDTD has been used to simulate the scattering spectrum of a single nanodisk embedded in a homogenous environment with refractive index = 1.5. The nanodisks are made of gold and they have height = 50 and radii = 50 and 80 nm. The scattering spectra are shown in Fig. S1, showing the resonance peaks are located at 682 and 779 nm respectively.

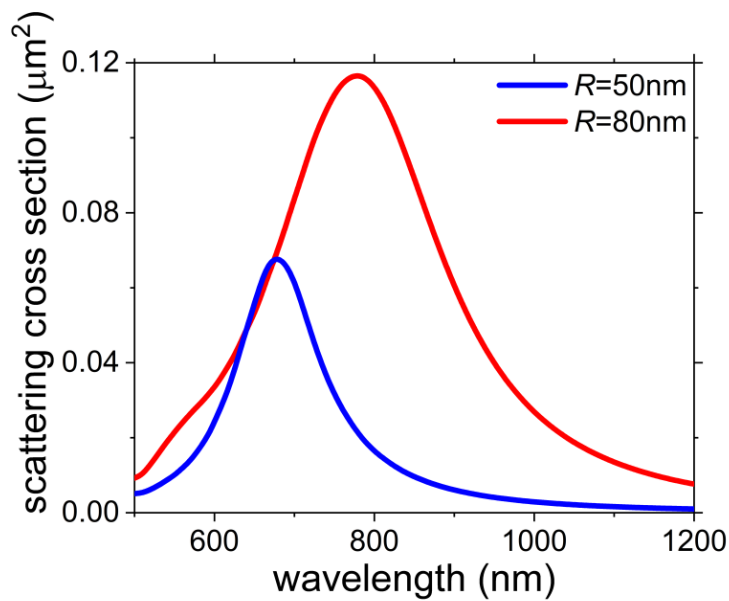

Fig. S1. The FDTD simulated scattering cross sections of nanodisks with $H = 50$ and $R = 50$ and 80 nm are plotted, showing the LSP resonances are located at 682 and 779 nm, respectively.

## B. DIAGONALIZATION OF DYNAMIC EQUATIONS

To diagonalize the bipartite system in Eq. (7), we solve for the eigenvalues and eigenvectors. Assuming the mode amplitudes have harmonic time dependence, such that $\frac{d}{dt}[a_1 \ a_2 \ a_3 \ a_4]^T = i\tilde{\omega}[a_1 \ a_2 \ a_3 \ a_4]^T$, where $\tilde{\omega} = \omega + i\Gamma/2$ is the complex angular frequency, we rewrite the homogeneous part of Eq. (7) as:

$$i\tilde{\omega}\begin{bmatrix} a_1 \\ a_2 \\ a_3 \\ a_4 \end{bmatrix} = i\begin{bmatrix} \tilde{\omega}_1 & \tilde{\Omega}_{12} & \Omega_{13} & \Omega_{14} \\ \tilde{\Omega}_{12} & \tilde{\omega}_1 & \Omega_{13} & \Omega_{14} \\ \Omega_{13} & \Omega_{13} & \tilde{\omega}_3 & 0 \\ \Omega_{14} & \Omega_{14} & 0 & \tilde{\omega}_3 \end{bmatrix}\begin{bmatrix} a_1 \\ a_2 \\ a_3 \\ a_4 \end{bmatrix}. \quad (B1)$$

The symmetry between $a_1$ and $a_2$ allowing us to reduce the Eq. (B1) by the transformation matrix

$$T_1 = \begin{bmatrix} 1/\sqrt{2} & 1/\sqrt{2} & 0 & 0 \\ -1/\sqrt{2} & 1/\sqrt{2} & 0 & 0 \\ 0 & 0 & 1 & 0 \\ 0 & 0 & 0 & 1 \end{bmatrix}, \text{ which transforms it into:}$$

$$i\tilde{\omega}\begin{bmatrix} a_1'' \\ a_2'' \\ a_3'' \\ a_4'' \end{bmatrix} = i\begin{bmatrix} \tilde{\omega}_1 - \tilde{\Omega}_{12} & 0 & 0 & 0 \\ 0 & \tilde{\omega}_1 + \tilde{\Omega}_{12} & \sqrt{2}\Omega_{13} & \sqrt{2}\Omega_{14} \\ 0 & \sqrt{2}\Omega_{13} & \tilde{\omega}_3 & 0 \\ 0 & \sqrt{2}\Omega_{14} & 0 & \tilde{\omega}_3 \end{bmatrix}\begin{bmatrix} a_1'' \\ a_2'' \\ a_3'' \\ a_4'' \end{bmatrix}, \quad (B2)$$

in which we can see $a_1''$ is decoupled and its corresponding eigenvalue is $\tilde{\omega}_1'' = \tilde{\omega}_1 - \tilde{\Omega}_{12} = \tilde{\omega}_1'$. The equation is now effectively reduced to a 3-by-3 matrix and we can then solve for the remaining eigenvalues by evaluating the determinant:

$$\begin{vmatrix} \tilde{\omega}_1 + \tilde{\Omega}_{12} - \tilde{\omega} & \sqrt{2}\Omega_{13} & \sqrt{2}\Omega_{14} \\ \sqrt{2}\Omega_{13} & \tilde{\omega}_3 - \tilde{\omega} & 0 \\ \sqrt{2}\Omega_{14} & 0 & \tilde{\omega}_3 - \tilde{\omega} \end{vmatrix} = 0, \quad (B3)$$

which gives the eigenvalues:

$$\tilde{\omega}_2' = \tilde{\omega}_3, \ \tilde{\omega}_{3,4}' = \frac{\tilde{\omega}_1 + \tilde{\Omega}_{12} + \tilde{\omega}_2}{2} \pm \sqrt{\left(\frac{\tilde{\omega}_1 + \tilde{\Omega}_{12} - \tilde{\omega}_2}{2}\right)^2 + 2\left(\Omega_{13}^2 + \Omega_{14}^2\right)}. \quad (B4)$$

We then construct the transformation matrix $T_2$ as:

$$T_2 = \begin{bmatrix} 1 & 0 & 0 & 0 \\ 0 & 0 & \beta_3(\tilde{\omega}_3 - \tilde{\omega}_3') & \beta_4(\tilde{\omega}_3 - \tilde{\omega}_4') \\ 0 & \beta_2\Omega_{14} & -\beta_3\Omega_{13} & -\beta_4\Omega_{13} \\ 0 & -\beta_2\Omega_{13} & -\beta_3\Omega_{14} & -\beta_4\Omega_{14} \end{bmatrix}, \quad (B5)$$

where $\beta_2 = \dfrac{1}{\sqrt{\Omega_{13}^2 + \Omega_{14}^2}}$, $\beta_3 = \dfrac{1}{\sqrt{(\tilde{\omega}_3 - \tilde{\omega}_3')^2 + \Omega_{13}^2 + \Omega_{14}^2}}$, and $\beta_4 = \dfrac{1}{\sqrt{(\tilde{\omega}_3 - \tilde{\omega}_4')^2 + \Omega_{13}^2 + \Omega_{14}^2}}$. Thus,

the overall transformation matrix can be written as $T = T_1 T_2$, and the eigenvectors are:

$$a_1' = \begin{bmatrix} -1 \\ 1 \\ 0 \\ 0 \end{bmatrix}, \quad a_2' = \begin{bmatrix} 0 \\ 0 \\ \Omega_{14} \\ -\Omega_{13} \end{bmatrix}, \text{ and } a_{3,4}' = \begin{bmatrix} \tilde{\omega}_3 - \tilde{\omega}_{3,4}' \\ \tilde{\omega}_3 - \tilde{\omega}_{3,4}' \\ -2\Omega_{13} \\ -2\Omega_{14} \end{bmatrix}. \quad (B6)$$

Noted that $T$ is invertible and thus $T^{-1} = T^T$. Therefore, Eq. (7), which has the form of,

$$\frac{d|a\rangle}{dt} = iH|a\rangle + K|S_+\rangle, \quad (B7)$$

can now be transformed as:

$$T^T \frac{d|a\rangle}{dt} = \frac{d|a'\rangle}{dt} = iT^T H T T^T |a\rangle + T^T K |S_+\rangle = iH'|a'\rangle + K'|S_+\rangle, \quad (B8)$$

where $H'$ is a diagonal matrix and is expressed as:

$$H' = \begin{bmatrix} \tilde{\omega}_1' & 0 & 0 & 0 \\ 0 & \tilde{\omega}_2' & 0 & 0 \\ 0 & 0 & \tilde{\omega}_3' & 0 \\ 0 & 0 & 0 & \tilde{\omega}_4' \end{bmatrix}. \quad (B9)$$

As a result, if $\omega_1 > \omega_3$, we assign $\omega_4' + i\dfrac{\Gamma_4'}{2}$ and $a_4'$ to be the nondegenerate (-1,0) SLR.

By comparing Eq. (2) and Eq. (B2), we learn that Eq. (2) is a special case of Eq. (7) and can be solved similarly. We substitute:

$$\tilde{\omega}_1 \to \tilde{\omega}_1 + \tilde{\Omega}_{12}, \quad \tilde{\omega}_2 \to \tilde{\omega}_3, \quad \Omega_{12} \to \sqrt{2}\Omega_{13}, \quad \Omega_{13} \to \sqrt{2}\Omega_{14} \quad (B10)$$

in Eq. (B2), and we follow the same procedure to solve for the eigenvalues and eigenvectors of Eq. (2):

$$\tilde{\omega}'_{1,2} = \frac{\tilde{\omega}_1 + \tilde{\omega}_2}{2} \pm \sqrt{\left(\frac{\tilde{\omega}_1 - \tilde{\omega}_2}{2}\right)^2 + \Omega_{12}^2 + \Omega_{13}^2}, \quad \tilde{\omega}'_3 = \tilde{\omega}_2, \tag{B11}$$

$$a'_{1,2} = \begin{bmatrix} \tilde{\omega}_2 - \tilde{\omega}'_{1,2} \\ -\Omega_{12} \\ -\Omega_{13} \end{bmatrix}, \text{ and } a'_3 = \begin{bmatrix} 0 \\ \Omega_{13} \\ -\Omega_{12} \end{bmatrix}. \tag{B12}$$

We can also write the transformation matrix T as:

$$T = \begin{bmatrix} \beta_1(\tilde{\omega}_2 - \tilde{\omega}'_1) & \beta_2(\tilde{\omega}_2 - \tilde{\omega}'_2) & 0 \\ -\beta_1\Omega_{12} & -\beta_2\Omega_{12} & \beta_3\Omega_{13} \\ -\beta_1\Omega_{13} & -\beta_2\Omega_{13} & -\beta_3\Omega_{12} \end{bmatrix}, \tag{B13}$$

where $\beta_1 = \dfrac{1}{\sqrt{(\tilde{\omega}_2 - \tilde{\omega}'_1)^2 + \Omega_{12}^2 + \Omega_{13}^2}}$, $\beta_2 = \dfrac{1}{\sqrt{(\tilde{\omega}_2 - \tilde{\omega}'_2)^2 + \Omega_{12}^2 + \Omega_{13}^2}}$, and $\beta_3 = \dfrac{1}{\sqrt{\Omega_{12}^2 + \Omega_{13}^2}}$.

Similarly, if $\omega_1 > \omega_{2,3}$, we assign $\tilde{\omega}'_2 = \omega'_2 + i\dfrac{\Gamma'_2}{2}$ and $a'_2$ to be the nondegenerate (-1,0) SLR, or $\omega_0 + i\dfrac{\Gamma_{tot}}{2}$ and $a$, as given in Eq. (3).

## C. TIME-DOMAIN METHOD WITH FDTD SIMULATIONS

Other than the CMT fitting, we also investigate the decay rates of a SLR single mode by using time-domain method. Briefly, a narrow bandwidth light pulse which barely covers the SLR mode of interest is used for excitation and the transient decay of the mode is then recorded in which its slope directly indicates $\Gamma_{tot} = \Gamma_{abs} + \Gamma_{rad}$. To differentiate $\Gamma_{rad}$ from $\Gamma_{tot}$, since the absorption is primarily due to the Ohmic dissipation arising from metal, we vary the imaginary part of the effective permittivity $\varepsilon''$ of gold from $\varepsilon''$ to $0.01\varepsilon''$ to sequentially reduce the absorption decay rate of the SLR and simulate the corresponding transients. As an illustration, at $\theta = 30°$ and $\varphi = 18°$ where (-1,0) SLR is excited at $\lambda = 921$ nm, the transients for different $\varepsilon''$ are shown in Fig. S2(a) and their decay slopes are extracted and then plotted in Fig. S2(b) against $\varepsilon''$. It is seen that the decay rate decreases with decreasing $\varepsilon''$ and the y-intercept indicates the decay rate where absorption is absent, thus yielding $\Gamma_{rad}$. $\Gamma_{abs}$ can then be determined by $\Gamma_{abs} = \Gamma_{tot} - \Gamma_{rad}$. All the time-domain results are overlaid in Fig. 5(a) and they agree very well with the CMT fittings with discrepancy within 10%.

## D. POLARIZATION ANGLE $\alpha$ FOR BEST ABSORPTION

To validate $\alpha$, which indicates the best in-coupling polarization of the incidence, we calculate the absorption of the SLR mode as a function of the incident polarization angle $\gamma$ defined with respect to the *s*-polarization. As an example, the absorption spectra of the (-1,0) SLR mode at $\lambda = 921$ nm taken at $\theta = 30°$ and $\varphi = 18°$ are plotted in Fig. S2(c) for several $\gamma$, showing the mode absorption at ~ 921 nm varies strongly with $\gamma$. It indicates there exists a certain polarization angle where the energy transfer from far-field to near-field is the best. We then plot the mode absorption at 921 nm as a function of $\gamma$ in Fig. S2(d). One sees the variation of the absorption exhibits a sinusoidal behavior and from the best fit we $\gamma = 27.2°$ yields the strongest absorption, which is consistent with the CMT fitting where $\alpha$ is determined to be 28.1°. By applying the same method to other $\varphi$, we determine all $\alpha$ and the results are also plotted in Fig. 5(b), showing the error is within 5%.

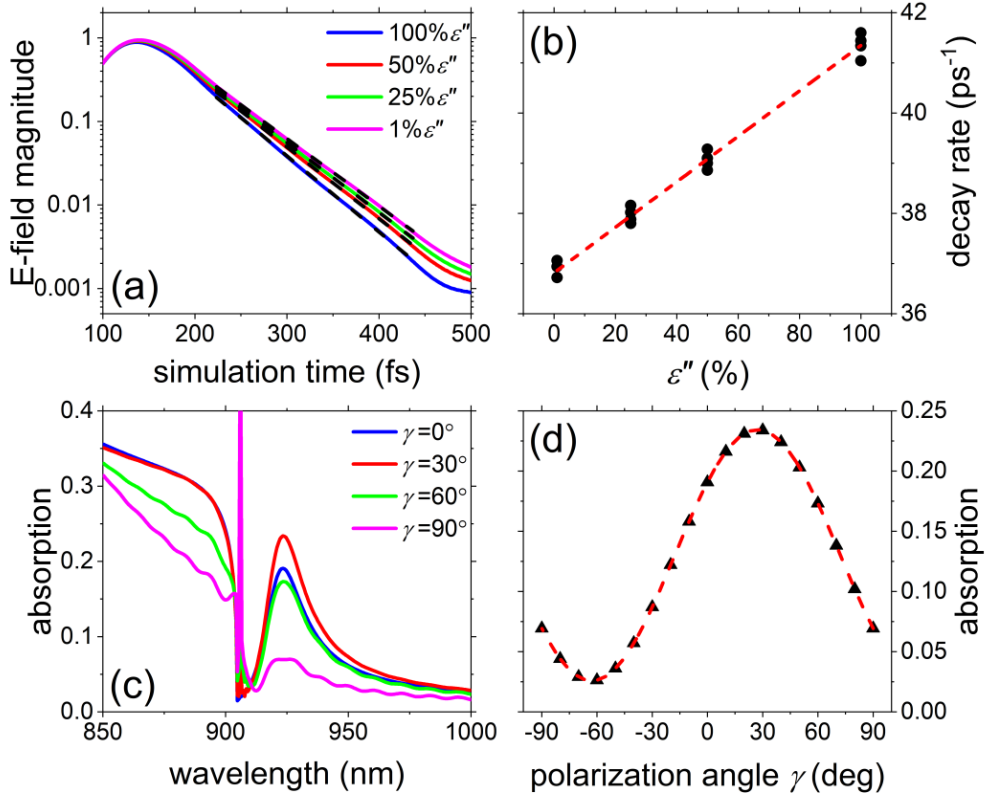

Fig. S2. At $\theta = 30°$ and $\varphi = 18°$, the (-1,0) SLR is excited at $\lambda = 921$ nm. (a) The transients of the electric field magnitude are plotted against the simulation time for different $\varepsilon''$. The dash lines are the linear fits for determining the decay rates. (b) The decay rates are extracted and then plotted against $\varepsilon''$. The dash line gives the y-intercept as $\Gamma_{rad}$. (c) The absorption spectra are taken for different $\gamma$, showing the mode absorption peaks at ~ 921 nm and it varies with $\gamma$. (d) The mode absorption at 921 nm is plotted as a function of $\gamma$ and fitted with a sinusoidal function, shown as the dash line. The strongest absorption is found at $\gamma = 27.2°$.

## E. EVALUATING THE QUALITY FACTOR OF SLR

The total decay rate of the monopartite SLR obtained from the diagonalization of Eq. (2) is given by $\Gamma_{tot} = \frac{\Gamma_1 + \Gamma_2}{2} - \text{Im}\left[\sqrt{\left(\omega_1 - \omega_2 + \frac{i}{2}(\Gamma_1 - \Gamma_2)\right)^2 + 4(\Omega_{12}^2 + \Omega_{13}^2)}\right]$ from Eq. (B11). Assume both $\Omega_{12}$ and $\Omega_{13}$ are real, we expand the second term by using $\text{Im}\sqrt{A + iB} = \sqrt{\frac{\sqrt{A^2 + B^2} - A}{2}}$ and find $\Gamma_{tot}$ to be:

$$\Gamma_{tot} = \frac{\Gamma_1 + \Gamma_2}{2} - \sqrt{\left(\frac{\Gamma_1 - \Gamma_2}{2}\right)^2 - \left[C - \sqrt{C^2 - 4\left(\frac{\Gamma_1 - \Gamma_2}{2}\right)^2 (\Omega_{12}^2 + \Omega_{13}^2)}\right]}, \tag{E1}$$

where $C = \frac{(\omega_1 - \omega_2)^2 + ((\Gamma_1 - \Gamma_2)/2)^2 + 4\Omega_{12}^2 + 4\Omega_{13}^2}{2}$. Since $C^2 > 4\left(\frac{\Gamma_1 - \Gamma_2}{2}\right)^2 (\Omega_{12}^2 + \Omega_{13}^2)$, we employ the binomial series to approximate $\sqrt{C^2 - 4\left(\frac{\Gamma_1 - \Gamma_2}{2}\right)^2 (\Omega_{12}^2 + \Omega_{13}^2)}$ and obtain:

$$\Gamma_{tot} = \Gamma_2 + \left(\frac{\Gamma_1 - \Gamma_2}{2}\right) - \sqrt{\left(\frac{\Gamma_1 - \Gamma_2}{2}\right)^2 - \frac{4((\Gamma_1 - \Gamma_2)/2)^2 (\Omega_{12}^2 + \Omega_{13}^2)}{(\omega_1 - \omega_2)^2 + ((\Gamma_1 - \Gamma_2)/2)^2 + 4\Omega_{12}^2 + 4\Omega_{13}^2} - \dots} \quad . \tag{E2}$$

Note that $\left(\frac{\Gamma_1 - \Gamma_2}{2}\right)^2 > \frac{4((\Gamma_1 - \Gamma_2)/2)^2 (\Omega_{12}^2 + \Omega_{13}^2)}{(\omega_1 - \omega_2)^2 + ((\Gamma_1 - \Gamma_2)/2)^2 + 4\Omega_{12}^2 + 4\Omega_{13}^2} + \dots$, so that $\Gamma_{tot}$ can be reduced to a simpler form as:

$$\Gamma_{tot} \approx \Gamma_2 + \frac{\Gamma_1 - \Gamma_2}{4}\left[\frac{4(\Omega_{12}^2 + \Omega_{13}^2)}{(\omega_1 - \omega_2)^2 + ((\Gamma_1 - \Gamma_2)/2)^2 + 4(\Omega_{12}^2 + \Omega_{13}^2)}\right]. \tag{E3}$$

Since RAs are dark modes, we assume $\Gamma_2 \ll \Gamma_1$ and further reduce $\Gamma_1 - \Gamma_2$ to $\Gamma_1$ and $\Gamma_{tot}$ becomes:

$$\Gamma_{tot} \approx \Gamma_2 + \frac{\Gamma_1}{4}\left[\frac{4(\Omega_{12}^2 + \Omega_{13}^2)}{(\omega_1 - \omega_2)^2 + (\Gamma_1/2)^2 + 4(\Omega_{12}^2 + \Omega_{13}^2)}\right]. \tag{E4}$$

Similarly, the resonant frequency of the monopartite SLR at $\omega_1 > \omega_2$ is given by $\omega_0 = \frac{\omega_1 + \omega_2}{2} - \text{Re}\left[\sqrt{\left(\frac{\omega_1 - \omega_2 + i(\Gamma_1 - \Gamma_2)/2}{2}\right)^2 + \Omega_{12}^2 + \Omega_{13}^2}\right]$ from Eq. (B11), and it can be approximated as:

$$\omega_0 \approx \omega_2 - (\omega_1 - \omega_2)\left[\frac{4(\Omega_{12}^2 + \Omega_{13}^2)}{(\omega_1 - \omega_2)^2 + (\Gamma_1/2)^2 - 4(\Omega_{12}^2 + \Omega_{13}^2)}\right], \quad (E5)$$

assuming $\frac{4(\Omega_{12}^2 + \Omega_{13}^2)}{(\omega_1 - \omega_2)^2 + (\Gamma_1/2)^2 - 4(\Omega_{12}^2 + \Omega_{13}^2)} \ll 1$. Finally, the Q factor, which is defined as $\omega_o/\Gamma_{tot}$, can be approximated as:

$$\frac{\omega_0}{\Gamma_{tot}} \approx \frac{4}{\Gamma_1}\left(\omega_2 \left(\frac{(\omega_1 - \omega_2)^2 + (\Gamma_1/2)^2}{4(\Omega_{12}^2 + \Omega_{13}^2)} + 2\right) - \omega_1\right), \quad (E6)$$

by letting $\Gamma_2$ in Eq. (E4) = 0.

The total decay rate of the bipartite SLR derived from CMT given in Eq. (B4) can be written as:

$$\Gamma_4' = \frac{\Gamma_1 + 2\Omega_{12}'' + \Gamma_3}{2} - \text{Im}\left[\sqrt{\left((\omega_1 + \Omega_{12}' - \omega_3) + i\left(\frac{\Gamma_1 + 2\Omega_{12}'' - \Gamma_3}{2}\right)\right)^2 + 8(\Omega_{13}^2 + \Omega_{14}^2)}\right]. \quad (E7)$$

When comparing with the total decay rate of the monopartite SLR $\Gamma_{tot}$, we find Eq. (E7) can be processed by using the substitutions in Eq. (B10). Thus, we derive a simpler form of $\Gamma_4'$ to be:

$$\Gamma_4' \approx \Gamma_3 + \frac{\Gamma_1 + 2\Omega_{12}''}{4}\left(\frac{(\omega_1 + \Omega_{12}' - \omega_3)^2 + (\Gamma_1/2 + \Omega_{12}'')^2}{8(\Omega_{13}^2 + \Omega_{14}^2)} + 1\right)^{-1}. \quad (E8)$$

Note that the conditions for the binomial expansions still hold after substitution.

Similarly, we can evaluate the resonant frequency of the bipartite SLR by substituting Eq. (B10) into Eq. (E5) and obtain:

$$\omega_4' \approx \omega_3 - (\omega_1 - \omega_3)\left[\frac{8(\Omega_{13}^2 + \Omega_{14}^2)}{(\omega_1 + \Omega_{12}' - \omega_3)^2 + (\Gamma_1/2 + \Omega_{12}'')^2 - 8(\Omega_{13}^2 + \Omega_{14}^2)}\right], \quad (E9)$$

under the condition $\frac{8(\Omega_{13}^2+\Omega_{14}^2)}{(\omega_1+\Omega_{12}'-\omega_3)^2+(\Gamma_1/2+\Omega_{12}'')^2-8(\Omega_{13}^2+\Omega_{14}^2)} \ll 1$. Thus, the Q factor of bipartite SLR, defined by $\omega_4'/\Gamma_4'$, can be approximated by:

$$\frac{\omega_4'}{\Gamma_4'} \approx \frac{4}{\Gamma_1+2\Omega_{12}''}\left(\omega_3\left(\frac{(\omega_1+\Omega_{12}'-\omega_3)^2+(\Gamma_1/2+\Omega_{12}'')^2}{8(\Omega_{13}^2+\Omega_{14}^2)}+2\right)-(\omega_1+\Omega_{12}')\right). \qquad (E10)$$